\documentclass[reprint,prl,aps,epsfig,longbibliography,superscriptaddress,twocolumn]{revtex4-2}
\usepackage{graphicx}
\usepackage[unicode=true,colorlinks=true]{hyperref}
\hypersetup{linkcolor=blue,citecolor=blue,urlcolor=blue}
\usepackage{dcolumn}
\usepackage{bm}
\makeatletter
\usepackage{color, colortbl}
\usepackage{mathtools, nccmath}
\usepackage{soul} 
\definecolor{Gray}{gray}{0.9}

\newcommand{\Rmnum}[1]{\expandafter\@slowromancap\romannumeral #1@}

\begin{document}
\title{Collapse of edge reconstruction in compressible Quantum Hall fluid within filling fraction range 2/3 to 1}

\begin{abstract}

The edge structure of a gate-defined compressible quantum Hall fluids in the filling fraction range 2/3 to 1 is studied using the three reconstructed $e^2/3h$ fractional edge modes of unity filling integer quantum Hall state. We find that the individually excited partially resolved $e^2/3h$ edge modes of the bulk state equilibrate completely even at higher magnetic field when passing through the gate defined compressible fluid with filling between 2/3 and 1. This result is unexpected because edge reconstruction at the smooth boundary is generally expected due to dominant incompressibility at filling 2/3 and 1/3. Recently such reconstructed edge mode has been reported for the compressible fluid in the filling fraction range 1/3 to 2/3. In contrary, equilibration of fractional edge modes in the compressible fluid within the filling fraction range 2/3 to 1 becomes faster with increasing magnetic field. This anomalous results will stimulate further investigations on edge structure in these complex many body systems.

\end{abstract}

\author{Suvankar Purkait} 
\email {suvankar.purkait@saha.ac.in}
\affiliation{Saha Institute of Nuclear Physics, 1/AF Bidhannagar, Kolkata 700 064, India}\affiliation{Homi Bhabha National Institute, Anushaktinagar, Mumbai 400094, India}
\author{Tanmay Maiti}
\affiliation{Saha Institute of Nuclear Physics, 1/AF Bidhannagar, Kolkata 700 064, India}
\affiliation{Homi Bhabha National Institute, Anushaktinagar, Mumbai 400094, India}
\affiliation{Department of Physics and Astronomy,Purdue University,West Lafayette,Indiana 47907,USA}
\author{Pooja Agarwal}
\affiliation{Saha Institute of Nuclear Physics, 1/AF Bidhannagar, Kolkata 700 064, India}
\affiliation{Homi Bhabha National Institute, Anushaktinagar, Mumbai 400094, India}
\affiliation{SPEC, CEA, CNRS, Université Paris-Saclay, CEA Saclay, 91191 Gif sur Yvette Cedex, France.}
\author{Suparna Sahoo}
\affiliation{Saha Institute of Nuclear Physics, 1/AF Bidhannagar, Kolkata 700 064, India}
\affiliation{Homi Bhabha National Institute, Anushaktinagar, Mumbai 400094, India}
\author{Giorgio Biasiol}
\affiliation{CNR-IOM – Istituto Officina dei Materiali, 34149 Trieste, Italy}
\author{Lucia Sorba}
\affiliation{NEST, Istituto Nanoscienze-CNR and Scuola Normale Superiore, Piazza San Silvestro 12, I-56127 Pisa, Italy}
\author{Biswajit Karmakar}
\email{ biswajit.karmakar@saha.ac.in}
\affiliation{Saha Institute of Nuclear Physics, 1/AF Bidhannagar, Kolkata 700 064, India}
\affiliation{Homi Bhabha National Institute, Anushaktinagar, Mumbai 400094, India}

 \maketitle
\maketitle

\subsection*{INTRODUCTION}
The topologically protected fractional edge modes at the boundary of the quantum Hall (QH) system transport quasi-particles \cite{Laughlin1983,Clark1988,Saminadayar1997,Elburg1998}. Hence, it is a very useful platform for understanding the characteristics of different quasi-particles \cite{de_Picciotto1997,Arovas1984,Heiblum2006,Radu2008,Faugno2019nabel, Bartolomei2020,Hashisaka2021,Biswas2022, Glidic2023,Manna2024prl}. In recent time, topologically protected fractional edge modes emerge as a promising platform for quasi-particle interferometry \cite{Halperin1984,McClure2012,Nakamura2019,Nakamura2020,Kundu2023,kim2024}, which has immense implication in quantum information processing \cite{Wen1991,Moore1991,Averin2001,Bonesteel2005, Kim2006,Nayak2008}. Therefore the studies of edge reconstruction and the equilibration of the reconstructed fractional edge modes are very crucial for quantum interferometric applications \cite{Bhattacharyya2019,Biswas2024} and other experiments like QH edge tunneling \cite{Wen1990,Chang1990,Wan2005,Hu2011,Varjas2013,Hashisaka2015, Cohen2023}, inter-edge interactions in confined geometry \cite{Fu2019, Nakamura2023prl, Yan2023} etc. Edge reconstruction alters the expected edge dynamics in different integer and fractional QH states. The reconstruction depends on various parameters, such as, the smoothness of the confining potential, strength of electron-electron interaction or magnetic field, and influence of disorder etc. Edge reconstruction  have been investigated extensively for different incompressible QH states
\cite{Chklovskii1992,Beenakker1990,Kouwenhoven1990,Wen1990prl,Joglekar2003,Wan2003,Yang2003, Paradiso2012,Zhang2014,Pascher2014, Khanna2021}.

Theoretical models and experimental observations of edge reconstruction for different QH fluid establish that multiple upstream and downstream modes might arise at the boundary \cite{MacDonald1990,Meir1994, Wan2002}. Further, these reconstructed edge modes might equilibrate among themselves and give the renormalized charge modes as well as charge neutral modes \cite{KFP1994,KF1995,Wang2013}. Therefore, the renormalized edge reconstruction depends on the experimental parameters and sample quality \cite{Meir1994}. One of the most studied edge reconstruction is the edge of hole-conjugate state in 2/3 FQH system \cite{Manna2024}, which is expected to host a downstream $e^2/h$ and a upstream $e^2/3h$ edge mode  under sharp confining potential \cite{MacDonald1990,Johnson1991}. Depending upon the interaction strength and propagation length, the modes get equilibrated and give rise to a downstream $2e^2/3h$ edge mode and a upstream charge neutral mode \cite{KFP1994,KF1995,Wang2013,Park2015,Protopopov2017,Nosiglia2018,Park2019, Spanslatt2021}. Experimental signature of such edge reconstruction is obtained at the graphene boundary \cite{Ckumar2018,Rkumar2022} and at the 2DES with sharp confining potential \cite{Bid2010,Melcer2022,Nakamura2023prl,Hashisaka2023}.

\begin{figure}[htbp]
\includegraphics[width=8.6 cm, height=6cm]{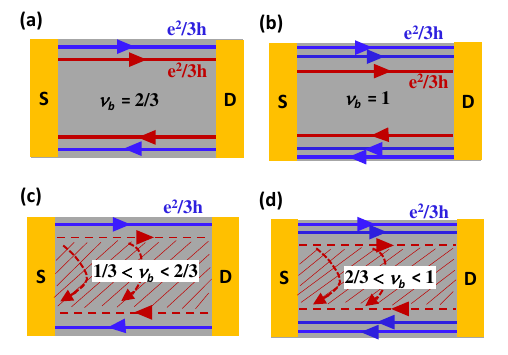}
\centering \caption[ ]
{\label{fig:edge reconstruction} (a) Schematic edge modes for 2/3 bulk filling fraction ($\nu_b$ where two downstream fractional modes with conductance $e^2/3h$ each are shown, other charge the neutral modes are not shown. (b) Schematics of reconstructed three downstream $e^2/3h$ fractional edge modes at $\nu_b$ = 1. (c) Schematic of edge reconstruction of compressible fluid with filling fraction between 1/3 and 2/3, where the outer $e^2/3h$ mode and inner bulk conduction channel is shown. (d) Schematic edge structure for compressible fluid with filling fraction $2/3 < \nu_b < 1$, where outer two $e^2/3h$ reconstructed modes and inner bulk conduction channel are expected.}
\end{figure}

At a very smooth confining potential, the edge of 2/3 FQH state is predicted to host more complex counter-propagating edge structure, which after equilibration give rise to two co-propagating $e^2/3h$ edge modes and neutral modes\cite{Meir1994,Wang2013}. Existance of such two co-propagating $e^2/3h$ edge modes is confirmed at the smooth boundary of 2DES at filling 2/3 \cite{sabo2017}. In subsequent experiment, two co-propagating edge modes with large equilibration length at filling 2/3 is demonstrated at high magnetic fields \cite{Maiti2020}. In the same experiment, three co-propagating $e^2/3h$ edge modes are observed for IQH state at unity filling fraction, where the innermost mode is seen to have large equilibration length of $\sim$ 800 $\mu$m at around 11 T magnetic field. Among these three modes, the outer two modes could not be resolved due to small equilibration length comparad to the large co-propagation length of 125 $\mu$m \cite{Maiti2020}.  However, previous experiment \cite{Bhattacharyya2019} and theoretical analysis \cite{Khanna2021} shows co-propagating $2e^2/3h$ and $e^2/3h$ edge modes for bulk unity filling fraction. These different results at filling fraction unity highlight the complexity of edge reconstruction, which depends on the the experimental conditions, sample quality and other sample parameters including strength of applied magnetic field. Notably, three co-propagating $e^2/3h$ edge modes at bulk filling unity, similar to our case (ref. \cite{Maiti2020}), have been observed in a 2DES previously \cite{Kouwenhoven1990} and Beenakker explained that the modes are originating from dominant incompressible strips of filling 1/3 and 2/3 at the boundary \cite{Beenakker1990}. We have adopted the model of Beenakker to analyzed the experimental data. The schematic edge structure for bulk filling fraction ($\nu_b$) 1 and 2/3 for our experimental parameters \cite{Maiti2020} are shown in Fig.\ref{fig:edge reconstruction}(a) and (b) respectively.

So far, edge reconstruction and edge equilibration are investigated for mostly the incompressible QH states, similar studies for compressible QH fluid is also important, since they might host fractional edge modes. Recently a reconstructed edge mode of conductance $e^2/3h$ is found at the boundary of the compressible QH fluid in the filling fraction range 1/3 to 2/3 \cite{Purkait2024}, as shown schematically in Fig. \ref{fig:edge reconstruction}(c). However, the edge structure of the compressible QH fluid in the filling fraction range 2/3 to 1 is not studied yet. Since co-propagating fractional edge modes are observed at filling fractions 2/3 as well as 1 in this sample, two downstream $e^2/3h$ fractional edge modes might be expected for compressible QH fluid in the filling fraction range $2/3 < \nu < 1$ in the same sample. The incompressible strips corresponding to 1/3 and 2/3 FQH fluid might be stable at the boundary of the compressible QH fluid with filling fraction 2/3 to 1 and give rise to two downstream $e^2/3h$ fractional edge modes \cite{Beenakker1990}. In line with this expectation, the edge structure for compressible QH fluid $2/3 < \nu < 1$ are schematically drawn in Fig.\ref{fig:edge reconstruction}(d). Such edge reconstruction has a direct impact on QH interferometry operating in this filling fraction region \cite{Bhattacharyya2019,Biswas2024}. In this work, our motivation is to experimentally verify presence of such expected reconstructed fractional edge modes (Fig.\ref{fig:edge reconstruction}(d)) in filling fraction range 2/3 to 1. 

In this article, we present experimental study on edge reconstruction of the gate defined compressible QH fluid for the filling fraction range 2/3 to 1. For this study we have utilized the experimental technique as reported in previous work \cite{Purkait2024}. We have selectively excited the partially resolved three $e^2/3h$ reconstructed fractional edge modes of integer QH state with bulk filling fraction $\nu_b =$ 1 connected to the gate-defined compressible fluid. Edge structure of the gate defined compressible FQH fluid with filling fraction in between 2/3 and 1 are probed by measuring the transmitted conductance of those excited fractional edge modes (of $\nu_b$ =1) through the gate-defined fluid.  Experimentally we observe that the excited $e^2/3h$ fractional edge modes fully equilibrate when passing through the gate defined compressible fluid with filling fraction in between 2/3 and 1. The results suggests that there might no reconstructed $e^2/3h$ fractional edge modes, or there might be other type of complex edge reconstruction, which is responsible for complete equilibration of $e^2/3h$ edge modes in this compressible filling fraction range. This result is in contrary to the general expectation of edge reconstruction \cite{Beenakker1990,Wan2002} and previous experimental observation of a fractional $e^2/3h$ edge mode at the boundary of the compressible QH liquid with filling fraction $1/3 < \nu < 2/3$ \cite{Purkait2024}. Therefore, our results indicate that the compressible FQH fluids in the filling fraction $2/3 < \nu < 1$ and $1/3 < \nu < 2/3$  are markedly different.

\begin{figure}[htbp]
\includegraphics[width=8.6 cm, height=10 cm]{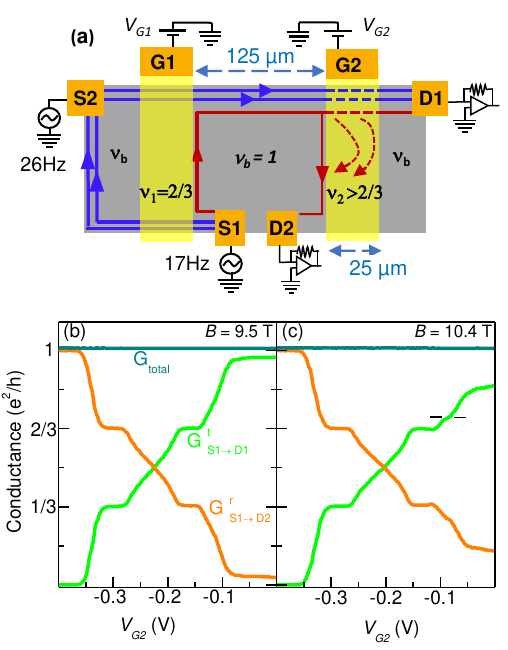}
\centering \caption[ ]
{\label{fig:device and characterization} (a) Schematics of topologically equivalent device structure along with the experimental setups. S1, S2, D1 and D2 are the Ohmic contacts for current injection and detection. G1 and G2 are top metal gates used for individually tuning the filling fractions $\nu_1$ and $\nu_2$ at bulk filling fraction $\nu_b$ = 1. Connections to the fractional edge modes are shown. The sources are excited with fixed voltages of 25.8 $\mu$ V and resulting currents at the drains are measured through  I-V preamplifiers. (b) and (c) Plots of two terminal conductance (TTC) (i.e. transmittance $G^{t}_{\rm S1 \rightarrow D1}$ and reflectance $G^{r}_{\rm S1 \rightarrow D2}$)  vs  G2 gate voltage ($V_{G2}$) at magnetic fields 9.5 T and 10.4 T respectively. The G2 gate characteristics are obtained by pinching-off the G1 gate completely and varrying the G2 gate voltage slowly. The sum of transmittance and reflectance gives the total conductance $G_{total}$, which is plotted in olive colored line. Robust conductance plateaus at $e^2/3h$ and $2e^2/3h$ in transmittance and reflectance are observed. In the high magnetic field data (plot (c)), an unidentified weak conductance structure is marked in between filling fraction 2/3 and 1.}
\end{figure}

\subsection*{EXPERIMENTAL SETUPS}

To probe the edge reconstruction in between 2/3 and 1 filling, a similar experimental techniques as in Ref. \cite{Maiti2020,Purkait2024} are utilized in a multi-terminal top gated 2DES device. The schematic device structure with measurement setup is shown in Fig.\ref{fig:device and characterization}(a). In the first set of experiment, the G1 gate is kept fixed at filling $\nu_2 = 2/3$ by applying proper $V_{G1}$ voltage with fixed bulk filling fraction $\nu_b$ = 1. In this condition, the outer two reconstructed $e^2/3h$ edge modes from S1 transmit through the G1 gate and reach contact S2, while the deflected inner 1/3 mode carry the excitation from S1 and propagates towards the detectors D1 and D2 \cite{Maiti2020}. The outer two modes emerge from S2 carrying the excitation with different frequency (26 Hz)  and propagates towards the detectors D1 and D2. The current at drain contacts are measured in different frequency by standard lock-in technique after amplifies with current-to-voltage pre-amplifiers. Such method of separately contacting the integer edge modes has been demonstrated in Ref. \cite{Karmakar2011}. In the current experiments, transmittance of the separately excited modes are studied by varying the filling fraction $\nu_2$ beneath G2 gate.

The low temperature injected carrier density and mobility of the 2DES are $n \sim 2.2\times10^{11} \ {\rm cm^{-2}}$ and $\mu \sim 4 \times10^6 \ {\rm cm^2/Vs}$ respectively at low temperatures. The measurements are done at dilution refrigerator with 7 mK base temperature. The device is initially characterized by measuring two-terminal magneto-conductance (2TMC) between the Ohmic contacts S1 and D2 with all other contacts open and grounding the two gates G1 and G2. The 2TMC shows various integer and fractional conductance plateaus down to bulk filling fractions $ \nu_b =$ 2/3 within 14 T of magnetic field \cite{Purkait2024}. The $\nu_b = 1$ conductance plateau in our sample is observed in the magnetic field range of 8 to 11 T. To characterize the gates (G1 and G2), transmitted conductance through the individual gates are measured by depleting the density below the gates from bulk filling fraction $ \nu_b =$1 by applying negative gate voltages. A similar characteristics for the two gates has been observed that confirms the uniformity of the sample (see appendix C) \cite{Maiti2020,Purkait2024}. The characteristics for G2 gate are shown in Fig. \ref{fig:device and characterization} (b) and (c) for magnetic fields $B$ = 9.5 T and 10.4 T respectively. Here the transmitted conductance $G^t_{S1 \rightarrow D1}$, plotted in green curve,  is measured between S1 and D1 with varying G2 gate voltage $V_{G2}$ when G1 gate is kept at fully pinched-off condition. At the same time the reflected conductance $G^r_{S1 \rightarrow D2}$, plotted in orange curve, is measured between S1 and D2. The FQH conductance plateaus at $e^2/3h$ and $2e^2/3h$ conductances in Fig.\ref{fig:device and characterization}(b) and (c) confirm the good quality and uniformity of the gate defined region. For the G2 gate filling fraction region in between $\nu_2 =$ 2/3 and 1, there is no fractional conductance plateau is observed in our sample for lower magnetic field (9.0 T, Fig\ref{fig:device and characterization} (b)). However, at higher magnetic fields (10.4 T), an unidentified weak conductance structure is observed (see also appendix A).

\begin{figure*}[htbp]
\includegraphics[width=18 cm, height=10 cm]{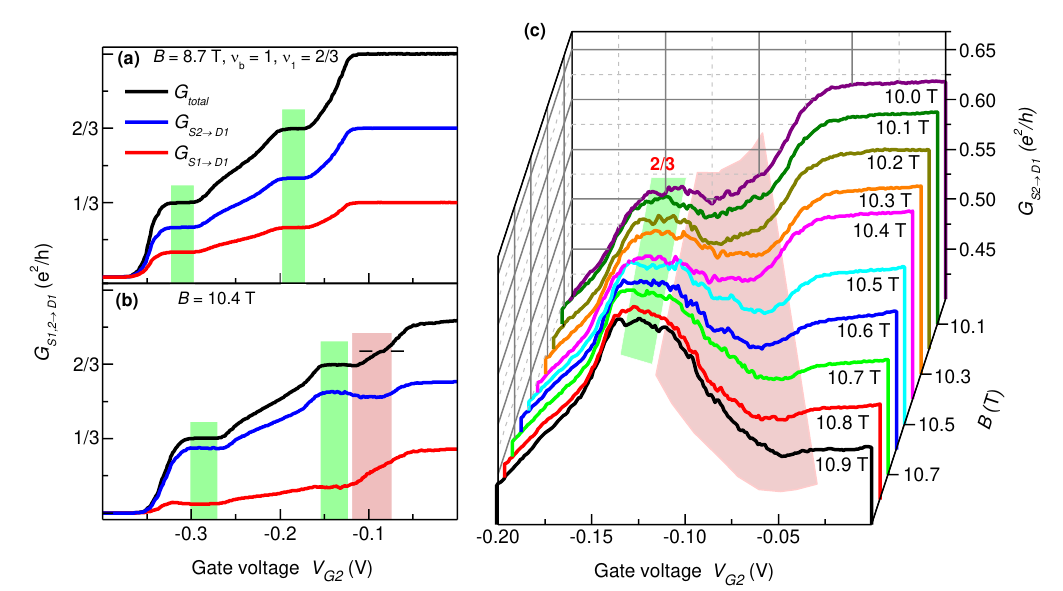}
\centering \caption[ ]
{\label{fig:g1 at 2/3 filling} (a) and (b) Two terminal conductances (TTCs) plotted against G2 gate voltage ($V_{G2}$) with fixed $\nu_1 = 2/3$ and  $\nu_b=1$ for magnetic fields $B$ = 8.7 T and 10.4 T respectively. Blue curves represent the transmittance $G^{~2/3,\nu_2}_{\rm S2 \rightarrow D1}{\rm (\nu=1, B)}$, red curves are for $G^{~2/3,\nu_2}_{\rm S1 \rightarrow D1}{\rm (\nu=1, B)}$. Black curve represents the total transmitted conductance $G_{total}$ at D1. Green shade indicates the $\nu_2$ = 1/3 and 2/3 FQH plateau regions. Brown shade highlights unexpected decrease of conductance $G^{~2/3,\nu_2}_{\rm S2 \rightarrow D1}{\rm (\nu=1, B)}$ above filling 2/3. This non-monotonic behavior is thye signature of edge equilibration. (c) Evolution of $G^{~2/3,\nu_2}_{\rm S2 \rightarrow D1}{\rm (\nu_b =1, B)}$ at different magnetic fields. At higher magnetic field,  $G^{~2/3,\nu_2}_{\rm S2 \rightarrow D1}$ conductance start to decrease for $G_{total} >$ 2/3 and form minima like structure (brown shaded region).}
\end{figure*}

Now we focus on the edge structure of the G2 gate defined compressible fractional quantum Hall fluid with filling fraction in the range $2/3 < \nu_2 <1$. For this study, we utilize the experimental setup as shown in Fig\ref{fig:device and characterization}(a). In this 2DES sample, the reconstructed edge modes of $\nu_b = 1$ integer QH system are well characterized, where three downstream fractional charge modes of conductance $e^2/3h$ each are obtained \cite{Maiti2020}, also drawn in Fig. \ref{fig:edge reconstruction}(a). The outer two reconstructed modes equilibrate with each other over the co-propagation length of 125 $\mu m$ for the experimental magnetic field range 8 to 11 T, while the inner mode fully equilibrates with the outer modes only at low magnetic field around 8 T. At higher magnetic fields, the inner mode does not fully equilibrate because of very high equilibration length of the order of 800 $\mu m$ \cite{Maiti2020}. Using these well characterized three $e^2/3h$ fractional modes, transmittance through the G2 gate defined FQH system is probed by individually exciting the modes as shown in Fig\ref{fig:device and characterization}(a). Here, the filling fraction $\nu_1$ beneath the gate G1 is set at 2/3 as shown in experimental setups in Fig\ref{fig:device and characterization}(a). Since 2/3 FQH fluid beneath the G1 gate have two co-propagating $e^2/3h$ modes, the outer two fractional edge modes, exited from source S2 with 25.8 $\mu$V 26 Hz excitation, will transmit through G1 gate. Similarly, the innermost edge mode excited from S1, carrying 25.8 $\mu$V 17 Hz excitation, will reflected by G1 gate and co-propagate 125 $\mu$m with the outer edge modes, which are excited from S2. During co-propagation, a fraction of current will be equilibrated among the edge modes depending upon the magnetic field dependent equilibration length \cite{Maiti2020}. The current from these edge modes will then either transmitted through the G2 gate and reach D1 or get reflected by the G2 gate and reach D2 depending upon the filling fraction $\nu_2$ beneath the G2 gate. The transmittance and reflectance of those individual fractional edge modes (which are excited from individual source contacts and carry different frequency excitation) are measured in different frequency windows by lock-in technique at contacts D1 and D2 (Fig. \ref{fig:device and characterization}(a)) .

\subsection{RESULTS}
For $\nu_1 = 2/3$ within $\nu_b = 1$, the two terminal conductances (TTCs) measured at drain contact D1 with varying G2 gate voltage $V_{G2}$ for magnetic fields 8.7 T and 10.4 T are plotted in Figure \ref{fig:g1 at 2/3 filling}(a) and (b). The measured TTCs are denoted as $G^{~\nu_1,\nu_2}_{\rm Si \rightarrow Dj}{\rm (\nu_b, B)}$, where i, j = 1, 2 are the indices of the source and detector contacts respectively. The red and blue curves in Figure \ref{fig:g1 at 2/3 filling}(a) and (b) represent the measured values of conductances $G^{~2/3,\nu_2}_{\rm S1 \rightarrow D1}{\rm (1, B)}$ and $G^{~2/3,\nu_2}_{\rm S2 \rightarrow D1}{\rm (1, B)}$ respectively. The sum of the above two conductances is the total conductance $G_{total}$ at D1 and is plotted in black color in Figure \ref{fig:g1 at 2/3 filling}(a) and (b). The curves of $G_{total}$ resembles the G2 gate characteristics as in Figure\ref{fig:device and characterization} (b) and (c), which confirms the resemblance of gate characteristics in independent measurements. The value of $G_{total}$ only depends on the G2 filling fraction $\nu_2$. The observed TTC plateau values (green shaded regions) evolve with magnetic field that is well understood in terms of equilibration properties of the reconstructed fractional edge modes \cite{Maiti2020}.

Note that, the results of Fig. \ref{fig:device and characterization}(c) and Fig. \ref{fig:g1 at 2/3 filling}(b) are from a completely different measurement conditions. The conductance in figure 2(c) is the G2 gate transmittance characteristics taken for G1 gate in pinch off condition ($\nu_1 = 0$). While the TTCs in Figure \ref{fig:g1 at 2/3 filling}(b) are the measured two terminal conductance when G1 gate is fixed at filling 2/3. Since the G1 gate is kept at filling $\nu_2 = 2/3$ for the later case, the outer two reconstructed $e^2/3h$ edge modes transmit beneath the G1 gate and reach contact S2 as shown in Fig. \ref{fig:device and characterization}(a), while the inner $e^2/3h$ mode is deflected and propagates towards the detectors D1 and D2. The inner $e^2/3h$ mode has larger equilibration length \cite{Maiti2020} at high magnetic field, hence, current from S1 to D1 is suppressed for $\nu_2 \le 2/3$ and $G^{~2/3,\nu_2}_{\rm S1 \rightarrow D1}$ is approaching to the limiting zero value \cite{Maiti2020}. This result can not be explained until considering three $e^2/3h$ edge modes. Interestingly, when the filling $\nu_2 \ge 2/3$, there is increase conductance $G^{~2/3,\nu_2}_{\rm S1 \rightarrow D1}$ because of enhanced equilibration of the edge modes and inner compressible region beneath the gate (see Fig. \ref{fig:device and characterization}(a)). This enhanced equilibration is the main focus of this article.

More elaboration of the results are as follows. The total transmittance depends on gate filling fraction as $G_{total} \sim \nu_2 e^2/h$ and hence the gate transmittance should evolve monotonically with the gate voltage $V_{G2}$. But surprisingly, the TTC $G^{~2/3,\nu_2}_{\rm S2 \rightarrow D1}{\rm (1, B)}$ value for $\nu_2 >$ 2/3 filling fraction starts to decrease gradually and then increases to form a minima like structure as marked with brown shade in Figure \ref{fig:g1 at 2/3 filling}(b). In this region, the conductance $G^{~2/3,\nu_2}_{\rm S1 \rightarrow D1}{\rm (1, B)}$ start to increase sharply after $\nu_2 > 2/3$, as seen from Figure \ref{fig:g1 at 2/3 filling}(b). 
These are the main results of this article. Evolution of TTC $G^{~2/3,\nu_2}_{\rm S1 \rightarrow D1}{\rm (1, B)}$ with magnetic field is presented in Figure \ref{fig:g1 at 2/3 filling}(c) (see also Fig. \ref{fig:additional data} of appendix section), where decreasing of $G^{~2/3,\nu_2}_{\rm S1 \rightarrow D1}{\rm (1, B)}$ for $\nu_2 > 2/3$ becomes prominent with increasing magnetic fields. This region of interest is marked in brown shade in fig. \ref{fig:g1 at 2/3 filling}. The unexpected reduction of $G^{~2/3,\nu_2}_{\rm S2 \rightarrow D1}{\rm (1, B)}$ indicates reduction of transmittance of the outer two modes (from S2) through G2 gate when filling fraction is $\nu_2 > 2/3$.
\begin{figure*}[htbp]
\includegraphics[width=18 cm, height=9 cm]{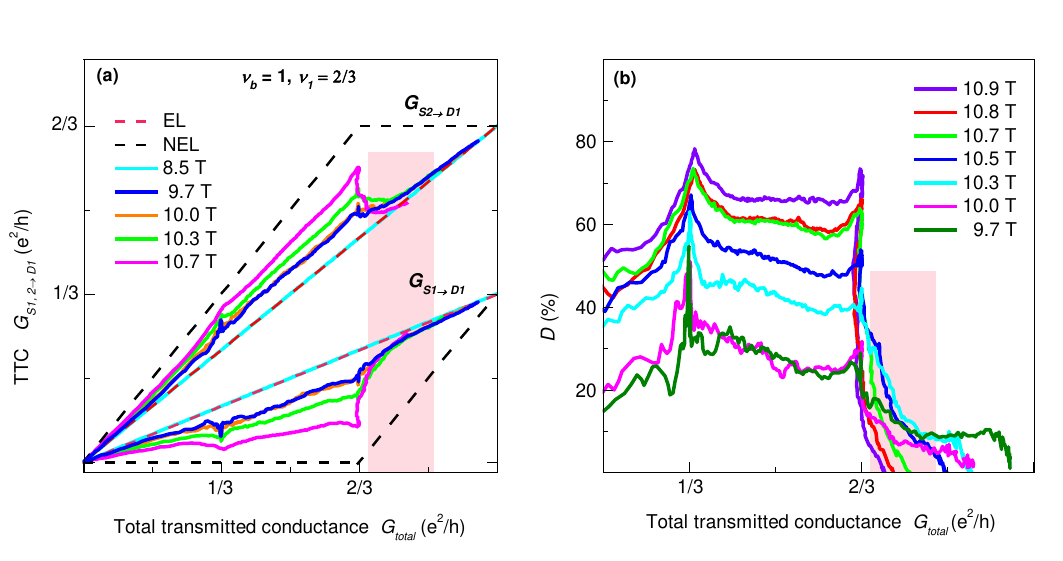}
\centering \caption[ ]
{\label{fig:Deviation from equilibration} (a) Plot of two terminal conductances (TTCs) vs total transmitted conductance ($G_{total}$) for different magnetic fields. The red dashed lines represent the full equilibration limits (EL) calculated from eqn.\ref{equation:1} and \ref{equation:2}. Black dashed lines represent the estimated non-equilibration limits (NEL) calculated from eqns.\ref{equation:3} to \ref{equation:6}. Upper bunch of curves are for $G^{~2/3,\nu_2}_{\rm S2 \rightarrow D1}{\rm (1, B)}$ and lower bunch represents $G^{~2/3,\nu_2}_{\rm S1 \rightarrow D1}{\rm (1, B)}$. The region of anomalous TTC behavior is indicated with shade. (b) Plot of deviation from equilibration $D$ (defined in eqn.\ref{equation:7}) vs $G_{total}$ for $G^{~2/3,\nu_2}_{\rm S2 \rightarrow D1}{\rm (1, B)}$ at different magnetic fields. The shaded region highlights the unexpected reduction of D value.}
\end{figure*}

To clearly visualize the unexpected reduction of $G^{~2/3,\nu_2}_{\rm S2 \rightarrow D1}$ for $\nu_2 > 2/3$, we plot the measured TTCs against the total transmitted conductance $G_{total}$ for different magnetic fields in Fig. \ref{fig:Deviation from equilibration}(a). Since, the total G2 transmission conductance depends on the filling fraction beneath the gate as $G_{total}\approx \nu_2  e^2/h$, the $G_{total}$ value approximately represents filling fraction $\nu_2$. The relation is exact for incompressible fractional states $\nu_2$ = 1/3 and 2/3. In Figure \ref{fig:Deviation from equilibration}(a), the upper bunch of the curves are TTCs for $G^{~2/3,\nu_2}_{\rm S2 \rightarrow D1}{\rm (1, B)}$ conductance and lower bunch represents $G^{~2/3,\nu_2}_{\rm S1 \rightarrow D1}{\rm (1, B)}$ conductances. Here, the TTCs are increasing quasi linearly with $G_{total}$ up to the value of 2/3. At lower magnetic fields, the outer two modes of $\nu_b = 1$ fully equilibrate with the inner mode during co-propagation and hence, we must observe the equilibrated values of the TTCs (as seen for 8.5 T). At the equilibration limit (EL) (when all three modes equilibrate), the TTCs can be expressed in terms of the product of two consecutive gate transmission probabilities, i.e.

\begin{equation} 
G^{~\nu_1,\nu_2}_{\rm S2 \rightarrow D1}{(\nu_b, B)} \mid_{EL} = \frac{(\nu_1 \times \nu_2)}
{\nu_b} \frac{e^2}{h} \ and
\label{equation:1}
\end{equation}
\begin{equation} 
G^{~\nu_1,\nu_2}_{\rm S1 \rightarrow D1}{(\nu_b, B)} \mid_{EL} = \frac{(\nu_b-\nu_1) \times \nu_2}{\nu_b} \frac{ e^2}{h}
\label{equation:2}
\end{equation}
The EL for both the TTCs (eqn.\ref{equation:1} and \ref{equation:2}) are plotted in Figure \ref{fig:Deviation from equilibration}(a) as red dashed lines. As expected, the plot of the measured TTCs at $B$ = 8.5 T (cyan lines) exactly follow the EL lines (red dashed lines) in Figure \ref{fig:Deviation from equilibration}(a).
At high magnetic field end of $\nu_b=1$ plateau, the inner most mode does not fully equilibrated with the outer two modes after co-propagation. Considering full equilibration of the outer two modes and complete non-equilibration limit (NEL) of the inner mode, the TTCs for $\nu_2 \le 2/3$ can be expressed as, 
\begin{equation} 
G^{~2/3,\nu_2}_{\rm S2 \rightarrow D1}{(\nu_b = 1, B)} \mid_{NEL} = \nu_2 \frac{ e^2}{h} \ and
\label{equation:3}
\end{equation}
\begin{equation} 
G^{~2/3,\nu_2}_{\rm S1 \rightarrow D1}{(\nu_b = 1, B)} \mid_{NEL} = 0.
\label{equation:4}
\end{equation}
For filling fraction range $2/3 < \nu_2 < 1$ the limiting values of the TTCs considering expected 
edge reconstruction can be denoted as,
\begin{equation} 
G^{~2/3,\nu_2}_{\rm S2 \rightarrow D1}{(\nu_b = 1,B)} \mid_{NEL} = \frac{2e^2}{3h}  \ and
\label{equation:5}
\end{equation}
\begin{equation} 
G^{~2/3,\nu_2}_{\rm S1 \rightarrow D1}{(\nu_b = 1,B)} \mid_{NEL} = (\nu_2 - 2/3) \frac{e^2}{h}.
\label{equation:6}
\end{equation}
The NEL of the TTCs for the whole filling fraction range (eqn.\ref{equation:3} to \ref{equation:6}) are plotted in black dashed lines in Figure \ref{fig:Deviation from equilibration}(a). With increasing magnetic field, the measured TTCs approach gradually towards the NEL curves for the filling fraction $\nu_2 \le 2/3$ as seen in Fig. \ref{fig:Deviation from equilibration}(a). These results prove the appropriateness of the adopted edge model to explain the NEL. Surprisingly, for $2/3 < \nu_2 < 1$  the TTCs do not follow the NEL (black dashed lines), instead they are reaching to full EL (red dasher lines) even at higher magnetic fields. Therefore, the results confirm the existence of strong equilibration process of fractional edge modes underneath the gate G2 for $2/3 < \nu_2 < 1$.

To quantify the amount of equilibration of the measured TTC (for $\nu_1$ = 2/3) at different magnetic fields, we define a physical quantity called deviation from equilibration ($D$) as,
\begin{equation} 
D \%  (\nu_2, B) = \frac{G^{~2/3,\nu_2}_{\rm S2 \rightarrow D1}(B) \mid_{measured} -\  G^{~2/3,\nu_2}_{\rm S2 \rightarrow D1} \mid_{EL}}{G^{~2/3,\nu_2}_{\rm S2 \rightarrow D1} \mid_{NEL} - \ G^{~2/3,\nu_2}_{\rm S2 \rightarrow D1} \mid_{EL}} \times 100\%
\label{equation:7}
\end{equation}
for the filling fraction range $0 < \nu_2 < 1$. The plot of $D$ versus $G_{total}$ in Figure \ref{fig:Deviation from equilibration}(b) shows that the value of $D$ increases with increasing magnetic field for the filling fraction range of $\nu_2 \le 2/3$ due to less equilibration of the inner mode with the outer two modes during co-propagation. The values of $D$ for a fixed magnetic field have peaks at $G_{total}$ = 1/3 and 2/3 because of adiabatic connections of the fractional edge modes for incompressible filling fractions. The value of $D$ reaches as high as 80 \% at the highest applied magnetic field 10.9 T because of lower equilibration. However, the value of $D$ is approaching full equilibration value ($D$ = 0) for $G_{total}$ above $2e^2/3h$ conductance. With increasing magnetic field, full equilibration ($D$ = 0) occurs at lower value of $G_{total}$. Therefore, in the filling fraction range $2/3 <\nu_2 < 1$, the fractional edge modes fully equilibrate below the gate.

\subsection{DISCUSSIONS}
At higher magnetic fields, the incompressible strip of 2/3 FQH state at the boundary separating the outer two modes from the inner compressible region for $\nu_2 > 2/3$ \cite{Beenakker1990,Chklovskii1992} is expected to be more pronounced, which should prevent equilibration between the edge modes with increasing magnetic field. However, the result in Figure \ref{fig:Deviation from equilibration}(b) shows opposite behavior.

It is important to note that our observation of edge mode equilibration in the gate defined FQH fluid is not arising from sample anomaly or inhomogeneity of the gate \cite{Purkait2024}. Equilibration of fractional edge modes in the gate defined FQH fluid of filling 2/3 to 1 at higher magnetic field is also observed in similar experiments by setting $\nu_1$ = 1/3 (see appendix). The additional data set confirms repeatability and reproducibility of enhanced equilibration at $\nu_2 > 2/3$.

In our previous work \cite{Purkait2024} with bulk filling 2/3, we have shown that the compressible fluid with filling $1/3 < \nu_2 < 2/3$ host the outer $e^2/3h$ reconstructed mode. If the disorders of localized states beneath the top gates (G2) cause enhanced equilibration, the compressible fluid ($1/3 < \nu_2 < 2/3$) must not host 1/3 edge mode. Therefore influence of disorders or localized states beneath the top gates is not noticeable in our studies. In the same sample, we have carried out the experiments in the compressible fluid $2/3 < \nu_2 < 1$. Hence influence of disorders or localized states beneath the top gates is not expected in the current study. Additionally if there is significant disorder, the gate characteristics should have hysteresis, while such hysteresis is not seen in our device(see appendix C). 

Presence of multiple $e^2/3h$ reconstructed edge modes at filling fraction 2/3 and 1 is well established. So conventionally edge reconstruction is expected in the filling fraction range 2/3 to 1. However, fractional edge modes fully equilibrate in the compressible fluid with the filling fraction 2/3 to 1. Therefore, our results indicate that the compressible FQH fluid in the filling fraction in between 2/3 and 1 might not support expected edge reconstruction and the compressible fluid is markedly different from the compressible QH liquid with filling fraction 1/3 to 2/3, which hosts a fractional edge mode \cite{Purkait2024}.

If one looks the results only in the experimental view point, low magnetic field data coincides with full equilibration line in figure \ref{fig:Deviation from equilibration}(a) and  \ref{fig:data for 1/3}(e). At higher magnetic fields, there is deviation from the equilibration limit towards non-equilibration limit, which proves expected enhancement of  equilibration length of fractional edge modes for $\nu_2 <2/3$ in figures figure \ref{fig:Deviation from equilibration}(a) and  \ref{fig:data for 1/3}(e). But, the enhanced equilibration above filling $\nu_2 > 2/3$ even at higher magnetic field is clearly seen. So, our interpretation does not depend on specific edge model.

The origin of the collapse of expected edge reconstruction even at high magnetic field is unclear. Notably, with increasing magnetic field the equilibration becomes faster as shown in fig 4(b). The result indicate that the collapse of expected edge reconstruction might be originating from enhanced correlation with increasing magnetic field or there might be more complex edge mode structure in this filling fraction region, which might be responsible for complete equilibration of co-propagating $e^2/3h$ modes. Our results will stimulate further theoretical and experimental investigations in this direction.

\subsection{CONCLUSIONS}
In conclusion, we have studied the edge structure of a gate-defined compressible QH fluid with filling fraction range 2/3 to 1 utilizing the individually excited resolved fractional edge modes of bulk unity filling integer QH state. We have found that the excited factional $e^2/3h$ edge modes equilibrate completely when passing through the gate-defined compressible fluid with filling fraction range 2/3 to 1, even at higher magnetic field where less equilibration is expected. The result suggest that the compressible QH fluid above filling fraction 2/3 does not support conventional edge reconstruction, in comparison the compressible QH fluid below filling fraction 2/3 hosts a fractional $e^2/3h$ edge mode.   \\

\section{APPENDIX}
\subsubsection{A. Additional data of edge equilibration  for $\nu_1$ = 2/3}
\begin{figure}[b!htp]
\includegraphics[width=8.6 cm, height=13.0 cm]{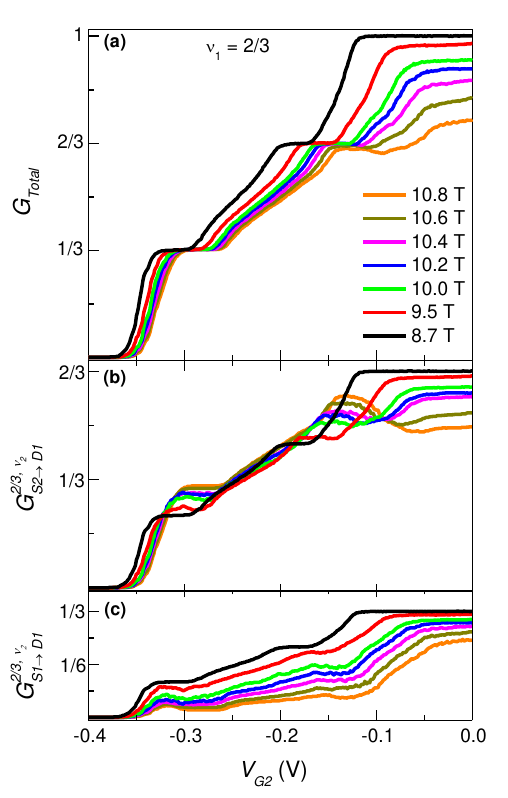}
\centering \caption[ ]
{\label{fig:additional data}Additional magnetic field dependent data of TTCs for $\nu_1$ = 2/3 and $\nu_b$ =  1. (a) Total conductance $G_{total}$ vs $V_{G2}$ for different magnetic fields. (b) Plots of $G^{2/3, \nu_2}_{\rm S2 \rightarrow D1}{\rm (\nu_b=1)}$ conductance versus $V_{G2}$.(c)Plots of $G^{2/3, \nu_2}_{\rm S1 \rightarrow D1}{\rm (\nu_b=1)}$ conductance vs $V_{G2}$. }
\end{figure}

Measured  TTCs at $\nu_1$ = 2/3 are presented in figure \ref{fig:additional data}(a), (b) and (c) for additional magnetic fields. Figure \ref{fig:additional data}(a) represents $G_{total}$, (b) represents $G^{2/3, \nu_2}_{\rm S2 \rightarrow D1}{\rm (\nu_b=1)}$ and (c) represents $G^{2/3, \nu_2}_{\rm S1 \rightarrow D1}{\rm (\nu_b=1)}$ curves. Here the curves are plotted on top of each other to clearly visualize the evolution of conductances. The unidentified conductance structure in between 2/3 and 1 filling is very weak and seen to evolve with magnetic fields. Therefore, no specific FQH state can be assigned to this conductance structure. In figure \ref{fig:additional data}(b), The $G_{\rm S2 \rightarrow D1}$ conductance value is seen to be reduced and becomes non-momotonous for the filling fraction region $\nu_2 > 2/3$. This reduction in $G_{\rm S2 \rightarrow D1}$ conductance is due to the equilibration of outer two modes with the inner compressible region beneath the G2 gate. In case of the full edge equilibration, the current from outer two modes will reduce, which results in the reduction of S2 to D1 transmittance.  The other consequence of equlibration is the increases transmission of inner most channel, which is seen as increment of the $G_{\rm S1 \rightarrow D1}$ conductance in figure \ref{fig:additional data}(c).

\subsubsection{B. Edge equilibration results for $\nu_1$ = 1/3}
\begin{figure*}[b!htp]
\includegraphics[width=18 cm, height=19 cm]{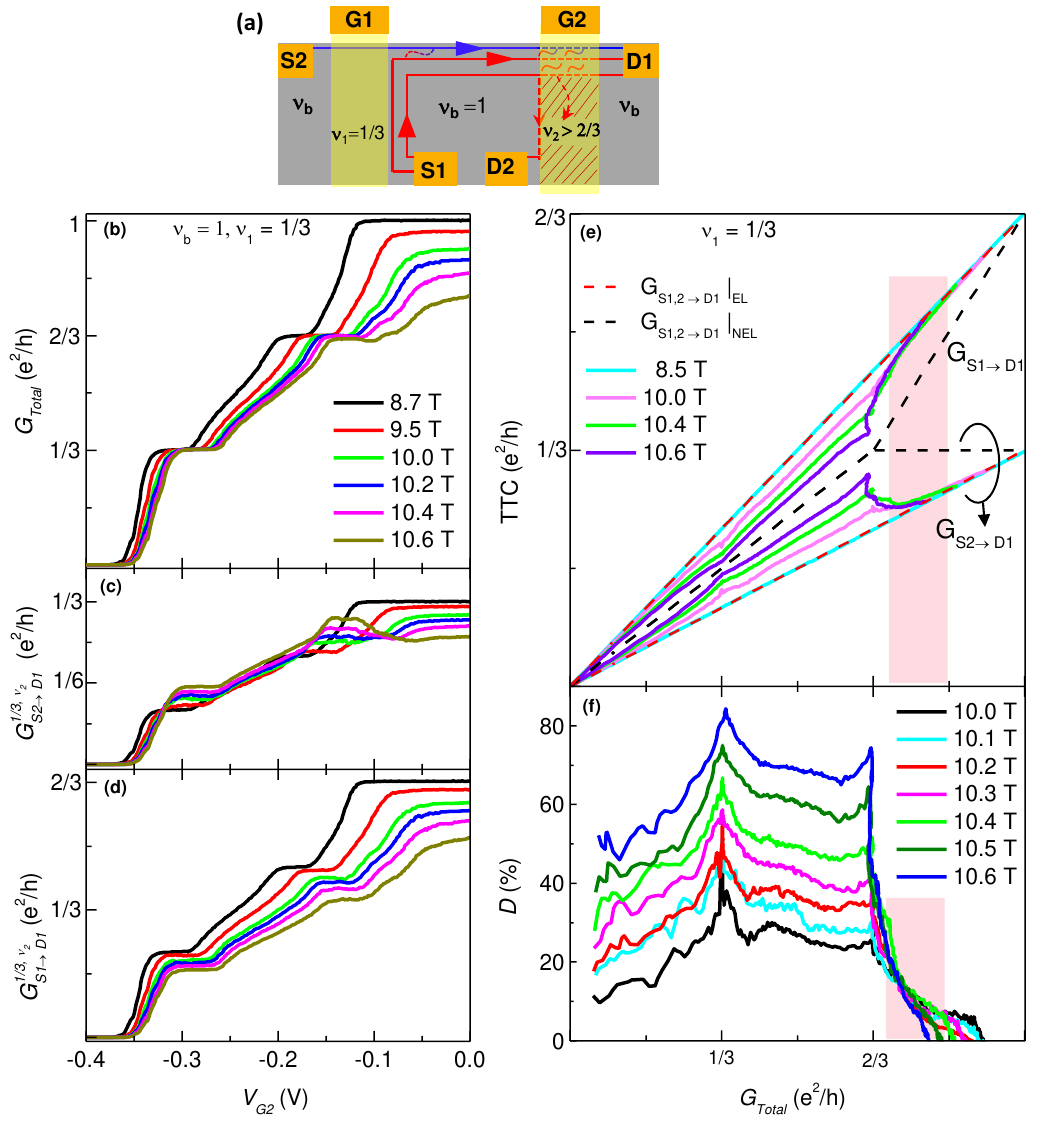}
\centering \caption[ ]
{\label{fig:data for 1/3} (a) Schematic setup for edge equilibration experiment at $\nu_1$ = 1/3 and $\nu_b$ = 1 along with edge connections. (b)-(d), Two terminal conductances (TTCs) plotted against G2 gate voltage ($V_{G2}$) with fixed $\nu_1 = 1/3$ and  $\nu_b=1$ for different magnetic fields. Total transmittance $G_{total}$ is plotted in (b), plots of $G^{1/3,\nu_2}_{\rm S2 \rightarrow D1}{\rm (\nu=1, B)}$ are shown in (c). Plots of $G^{1/3,\nu_2}_{\rm S1 \rightarrow D1}{\rm (\nu=1, B)}$ are presented in (d). (e) Plots of $G^{1/3,\nu_2}_{\rm S2 \rightarrow D1}$ and $G^{1/3,\nu_2}_{\rm S1 \rightarrow D1}$ versus $G_{total}$. The region of unexpected edge equilibration is highlighted with shade. (f) Plots of calculated deviation from equilibration, $D$ vs $G_{total}$. Note the sharp drop of D in the shaded region.}
\end{figure*}

The finding is corroborated in another edge mode excitation configuration by setting the G1 gate filling fraction at $\nu_1$ =1/3 within bulk filling $\nu_b$ = 1. In this configuration the outermost mode is excited from S2 source. The middle and innermost modes are excited from source S1. During co-propagation in between two gates, the outer two modes will fully equilibrate, while the innermost mode will remain resolvable at higher magnetic fields. The schematic edge mode connections in this configuration are shown in figure \ref{fig:data for 1/3}(a). Corresponding measured conductance are plotted in Figure \ref{fig:data for 1/3} (b)-(d). The limiting values of the conductances can be estimated as before and they will be different from the previous case with $\nu_1$ = 2/3.

In case of full equilibration of the three edge modes, the limiting values of TTCs can be given by equations \ref{equation:1} and \ref{equation:2} as before and is plotted in black dashed line in Figure \ref{fig:data for 1/3}(e). At higher magnetic field, the innermost mode will have less equilibration with the outer two modes, which are fully equilibrated though in our experimental conditions \cite{Maiti2020}. In the high magnetic fields, we can estimate the non-equilibration limit (NEL) of the conductances.  The estimated NEL values of the TTCs, plotted in red dashed line in Figure \ref{fig:data for 1/3}(e) for $\nu_2 \le 2/3$ can be written as,
\begin{equation} 
G^{~1/3,\nu_2}_{\rm S2 \rightarrow D1}{(1,B)} \mid_{NEL} = \frac{ \nu_2e^2}{2h}  \ and
\label{equation:8}
\end{equation}
\begin{equation} 
G^{~1/3,\nu_2}_{\rm S1 \rightarrow D1}{(1,B)} \mid_{NEL} = \frac{ \nu_2e^2}{2h}.
\label{equation:9}
\end{equation}
Similarly, for filling fraction range $2/3 \le \nu_2 \le 1$, 
\begin{equation} 
G^{~1/3,\nu_2}_{\rm S2 \rightarrow D1}{(1,B)} \mid_{NEL} = \frac{e^2}{3h} \  and
\label{equation:10}
\end{equation}
\begin{equation} 
G^{~1/3,\nu_2}_{\rm S1 \rightarrow D1}{(1,B)} \mid_{NEL} = (\nu_2 - 1/3) \frac{e^2}{h}.
\label{equation:11}
\end{equation}

Here also, the TTCs are increasing quasi linearly in Figure \ref{fig:data for 1/3}(e) with filling $G_{total}$. The TTCs increases towards the NEL line (eqn. 8-11, plotted in black dashed lines) with increasing magnetic field upto the $G_{total}$ = 2/3 and then start approaching to the EL (eqn. 1 and 2, plotted in red dashed lines)  for $2/3 < \nu_2 < 1$. The quantity $D$ is also calculated for $\nu_1$=1/3 using the equation \ref{equation:7} and plotted in fig. \ref{fig:data for 1/3}(f), which clearly shows the complete equilibration of edge modes for $\nu_2 > $2/3 filling fraction. Therefore, for filling fraction $\nu_2 > 2/3$, the complete equilibration of expected edge modes is evident as stated in the main text. \\

\subsection{C. Individual gate characteristics and repeatability}
To check the quality of the device, we performed the individual gate voltage 
\begin{figure}[b!htp]
\includegraphics[width=8.6 cm, height=10 cm]{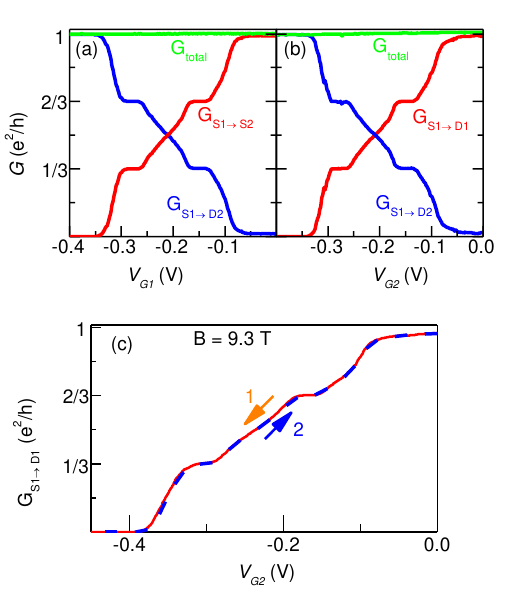}
\centering \caption[ ]
{\label{fig:gate characteristics and hysteresis} (a) Individual gate characteristics at B = 8.85 T of gates G1 when the G2 gate in fully pinched-off.  (b) Gate characteristics of gates G2 obtained with the G2 gate in fully pinched-off condition. (c) Back and forth gate sweep data for G2 gate at B= 9.3 T. Plot shows no hysteresis in gate sweeps.}
\end{figure}
sweeps at different magnetic fields. Representative gate sweeps for both the gates G1 and G2 are plotted in figure \ref{fig:gate characteristics and hysteresis}(a) and (b) at B = 8.85 T. The plots show that the individual gate characteristics are similar, which confirms symmetry and uniformity the sample. In figure \ref{fig:gate characteristics and hysteresis}(c), we have presented the back and forth gate voltage sweeps for G2 gate, which does not show hysteresis due to the localized states.

\subsection{ACKNOWLEDGMENTS}
Authors thank Sourin Das,  G J Sreejith, Ankur Das and Krishanu Roychowdhury for insightful discussions and valuable suggestions.

\bibliography{edgeequilibration}

\end{document}